\documentclass[prl,twocolumn,floatfix]{revtex4}

\usepackage{epsf,amsmath,amsfonts} 
\usepackage{graphicx}
\usepackage{setspace}
\DeclareGraphicsExtensions{.ps}

\usepackage{amsmath,amssymb}
\usepackage{latexsym}
\usepackage{fancyhdr}

\usepackage{dcolumn}   % Align table columns on decimalpoint
\usepackage{bm}        % bold math
\usepackage{subfigure}

\newcommand{\pt}{\partial}

\begin{document}

\title{Dynamic buckling and fragmentation in brittle rods}

\author{J. R. Gladden$^\dagger$}

\author{N. Z. Handzy$^\dagger$}

\author{A. Belmonte$^\dagger$}

\author{E. Villermaux$^\ddagger$}
\thanks{Also at: Institut Universitaire de France.}

\affiliation{$^\dagger$The W.~G.~Pritchard Laboratories, Department of Mathematics,
The Pennsylvania State University, University Park, PA 16802, USA\\
$^\ddagger$Institut de Recherche sur les Ph\'enom\`enes Hors Equilibre, Universit\'e de Provence\\ 49, rue Fr\'ed\'eric Joliot-Curie 13384 Marseille Cedex 13, FRANCE}

\date{\today}

%\date{\today; submitted January 10, 2003}

%\footnote{Also at: Institut Universitaire de France}

%(REVISED VERSION)

%(\today, at \currenttime \quad - \quad \jobname .tex)

\begin{abstract}
We present experiments on the dynamic buckling and fragmentation of slender rods axially impacted by a projectile.  By combining the results of Saint-Venant and elastic beam theory, we derive a preferred wavelength $\lambda$ for the buckling instability, and experimentally verify the resulting scaling law for a range of materials including teflon, dry pasta, glass, and steel.  For brittle materials, buckling leads to the fragmentation of the rod.  Measured fragment length distributions show two clear peaks near $\lambda/2$ and $\lambda/4$. The non-monotonic nature of the distributions reflect the influence of the deterministic buckling process on the more random fragmentation processes.
\end{abstract}

\maketitle

%%%%%%%%%%%%%%%%%%%%%%%%%%%%%%%%%%%%%%%%%%%%%%%%%%%%%%%%%%%%%

Long, thin supports are ubiquitous in natural and engineered load bearing structures, from spider legs to the steel struts of a skyscraper \cite{thomp, vogel}. A single rod will buckle if too much force is applied along its axis, which can lead to the catastrophic failure of the structure. The buckling instability is seen at all sizes, from pole vaulting \cite{ekevad97} to protein microtubules confined in vesicles \cite{elbaum96} and carbon nanotube atomic force microscope probes \cite{falvo97}. While the classic {\it Euler buckling} of a rod is due to a static axial load \cite{love}, a different physical process occurs when the stress is applied suddenly, as during impact \cite{jones, goldsmith, stronge}.
In this Letter, we show that this ``dynamic buckling'' \cite{lindberg} obeys a simple scaling law, which we derive by combining the approach of Saint-Venant with the classical theory of elastic rods \cite{love,jones}. For brittle rods, buckling often leads to breaking, for which we find the distribution of fragment lengths displays a unique non-monotonic shape reflecting the primary buckling instability.

%%%%%%%%%%%%%%%%%%%%%%%%%%%%%%%%%%%%%%%%%%%%%%%%%%%%%%%%%%%%%
\begin{figure}[t]
\begin{center}
\includegraphics[width=7.5cm]{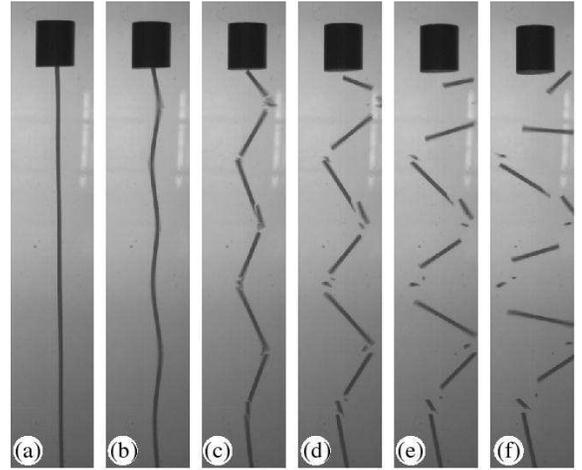}
\caption{\label{pasta} The dynamic buckling and fragmentation
of dry pasta ($d = 1.9$ mm, $L = 24$ cm)  just after the impact of an aluminum projectile at velocity $U_0 = $ 3.5 m/s (interval between pictures: 236 $\mu$s).}
\end{center}
\end{figure}
%%%%%%%%%%%%%%%%%%%%%%%%%%%%%%%%%%%%%%%%%%%%%%%%%%%%%%%

The process of dynamic buckling and subsequent fragmentation is illustrated in Fig.~\ref{pasta}, in which a falling weight strikes an upright brittle rod (dry pasta). Within a fraction of a millisecond after impact, a sinusoidal perturbation appears (Fig.~\ref{pasta}b), much different than the half-wavelength seen in Euler buckling. A few tenths of a millisecond later, the pasta has buckled appreciably and begins to shatter (Fig.~\ref{pasta}c). This imparts angular momentum of alternating signs to the fragments, which rotate and scatter (Fig.~\ref{pasta}d-f).Our experimental setup consists of a simple metal holder for the rod, and a pneumatic cannon in which a pressure reservoir at 60 psi delivers an impulse to a steel cylindrical projectile (1.46 cm diameter, 24.9 or 10.0 g) held at the end of a 1.5 m acrylic tube with two magnets. The holder rests on a steel plate in a sandbox which serves as a shock absorber to stop the projectile. The buckling and fragmentation dynamics were imaged by a high speed digital video camera (Phantom v5.0), capable of capturing up to 62,000 frames per second.
The speed of the projectile was also measured just before impact using the video system. The materials used included dry pasta ($\rho = 1.5$ g/cm$^3$, $d =$ 1.1 mm and 1.9 mm) \cite{pasta},
borosilicate glass ($\rho = 2.4$ g/cm$^3$, $d =$ 2.0 mm),
type 303 stainless steel ($\rho = 7.9$ g/cm$^3$, $d =$ 1.6 mm),
and teflon (PTFE) ($\rho = 2.2$ g/cm$^3$, $d =$ 2.0 mm),
all with circular cross-sections and lengths ranging from 14-29 cm. We also used strips of plastic and paper, which had rectangular cross-sections. The Young's modulus ($E$) was measured for the pasta by a clamped beam resonance technique \cite{jones}, averaging the fundamental frequency for five different pieces; we find $E \simeq 2.9$ GPa. Standard tabulated values of $E$ were used for the other materials: 62 GPa (glass), 200 GPa (stainless steel), and 0.5 GPa (PTFE).

The first manifestation of dynamic buckling is the onset of undulations with a well-defined wavelength $\lambda$. Our initial experiments involved simply dropping a metal weight from a given height, so that the impact velocity was fixed; the material, length, and thickness of the rod varied. We find that $\lambda$ is strongly dependent on the smallest dimension of the rod (its thickness or diameter $d$), and apparently independent of the length $L$; in contrast for Euler buckling, $\lambda = 2L$. For a fixed impact velocity $U_0 = 3.5$ m/s, measurement of the initial sinusoidal perturbation for a wide variety of materials (including paper, plastic, and glass) indicates that $\lambda \sim d$ as shown in Fig.~\ref{speed}a, seemingly independent of the elastic properties of each material. A more systematic study, however, shows that the material properties do indeed play a role. By varying the impact speed $U_0$ from 0.7 to 30 m/s using the pneumatic cannon, and widening our study to include materials with high and low sound speeds (stainless steel, $c = 5020$ m/s, and teflon, $c = 470$ m/s), we find that $\lambda \sim U_0^{-1/2}$, with a prefactor which varies with the material (see Fig.~\ref{speed}b).

%%%%%%%%%%%%%%%%%%%%%%%%%%%%%%%%%%%%%%%%%%%%%%%%%%%%%%%%
\begin{figure}[t]
\begin{center}
\includegraphics[width=4.5cm]{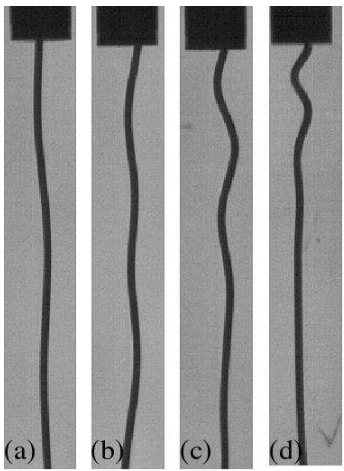}
\caption{Buckled shape of a cylindrical PTFE teflon rod ($d$=2.0 mm, $L$=14cm), shown $\sim250 \mu$s after impact by a steel projectile with speeds: (a) 0.7 m/s , (b) 4.6 m/s, (c) 11.2 m/s, and (d) 26.0 m/s.}
\label{allmat}
\end{center}
\end{figure}
%%%%%%%%%%%%%%%%%%%%%%%%%%%%%%%%%%%%%%%%%%%%%%%%%%%%%%%

The sequence of events leading to the buckling of the rod begins just after impact, as a stress wave penetrates into the material at the speed of sound. In the compressed region, the rod becomes elastically unstable and buckles due to the axial stress. We describe the destabilization by recalling the dynamical equation for the lateral displacement $\xi(x,t)$ of a thin straight elastic beam \cite{jones,love,goldsmith,lindberg} under an applied longitudinal force $F(x,t)$
\begin{equation} 
\rho A\frac{\pt^2 \xi}{\pt t^2} +\frac{\pt}{\pt x}\left (F(x,t)\frac{\pt \xi}{\pt x}\right) + EI\frac{\pt^4 \xi}{\pt x^4}=0
\label{pde}
\end{equation}
where $A$ is the cross-sectional area, $E$ is the Young's modulus, and $I$ is the moment of inertia of the area ($I=\pi d^4/64$ for a circular cross-section). Standard normal mode stability analysis of Eq.~(\ref{pde}) with a constant force $F(x,t)\equiv F_0$ yields the fastest growing unstable mode, with a wavelength $\lambda$ and associated growth rate $\tau_{\rm b}^{-1}$ given by
\begin{equation}\frac{\lambda}{d} = \frac{\pi^{3/2}}{2\sqrt{2}} \sqrt{\frac{E d^2}{F_0}} \qquad {\rm and}\qquad \tau_{\rm b}^{-1} = \frac{8 F_0}{\pi \rho c d^3},
\label{laws}
\end{equation}
where $c=\sqrt{E/\rho}$ is the speed of sound. 

%%%%%%%%%%%%%%%%%%%%%%%%%%%%%%%%%%%%%%%%%%%%%%%%%%%%%%%%
\begin{figure}[t]
\begin{center}
\includegraphics[width=8cm]{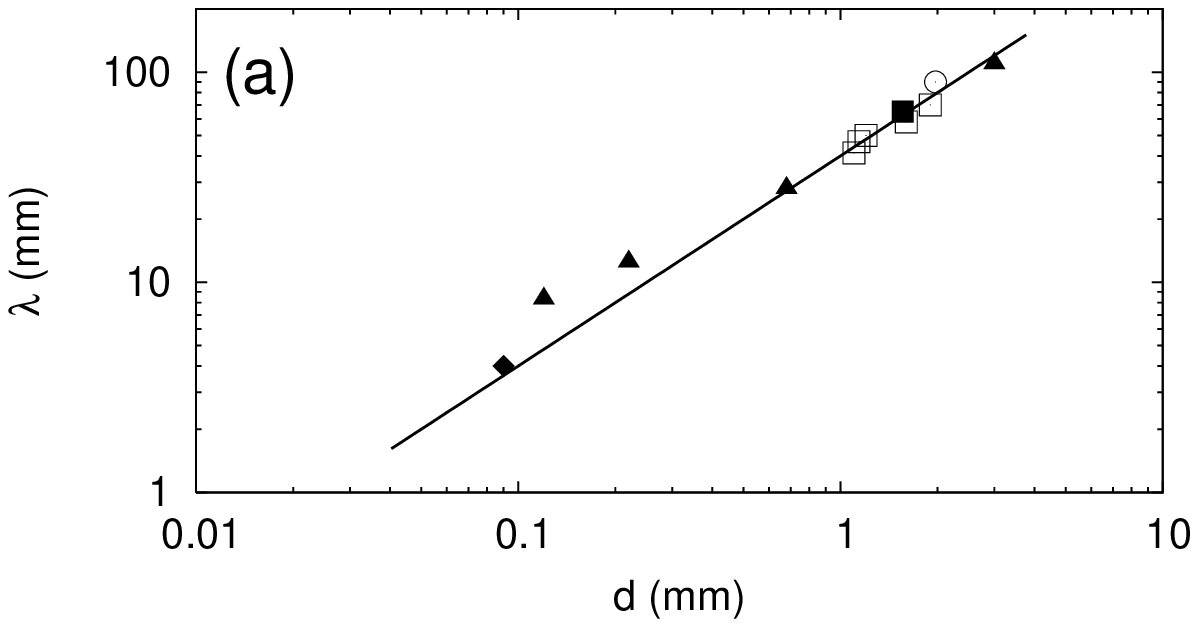}
\includegraphics[width=8cm]{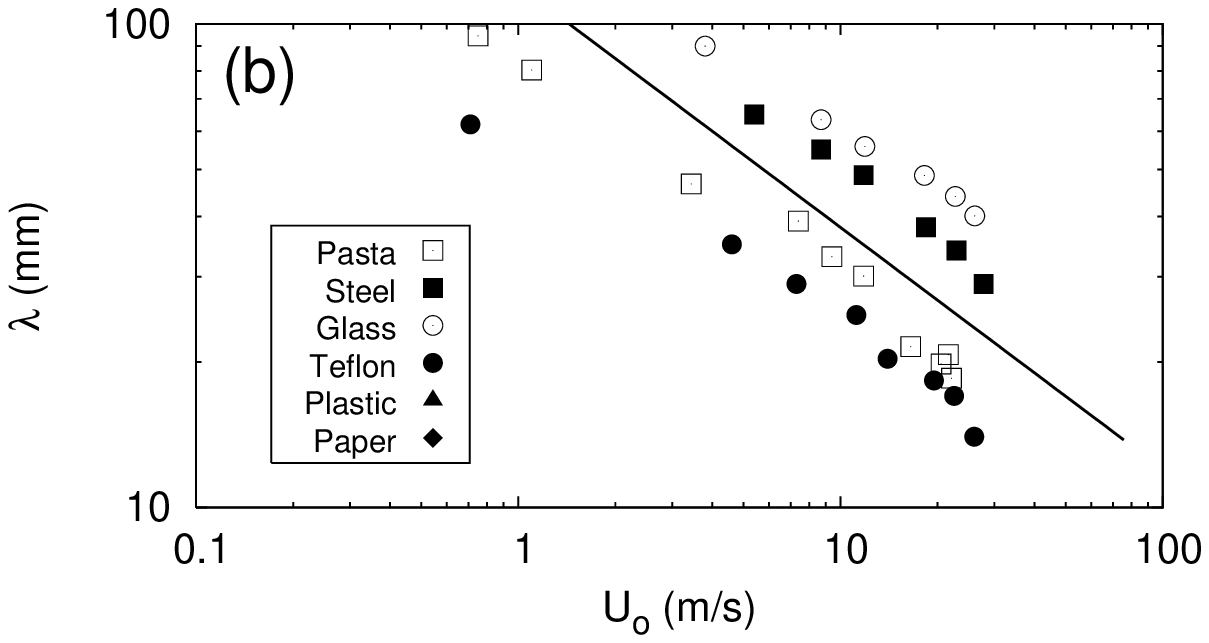}
\caption{The measured dynamic buckling wavelength for various materials versus (a) smallest dimension and (b) impact speed. The line in (a) corresponds to the scaling $\lambda \sim d$ and in (b) corresponds to the scaling $\lambda \sim U_0^{-1/2}$.}
\label{speed}
\end{center}
\end{figure}
%%%%%%%%%%%%%%%%%%%%%%%%%%%%%%%%%%%%%%%%%%%%%

The instability is driven by the applied force ($F_0$), which is the elastic response of the material to impact. In order to proceed we need to obtain this force from the initial momentum of the projectile and the properties of the rod.  The problem of stress in a rod fixed at one end ($x=L$) and struck longitudinally at the other ($x=0$) by a mass $M$ with initial speed $U_0$ was solved by Saint-Venant in 1883 \cite{love}. In the linear elasticity limit, the longitudinal displacement $w(x,t)$ obeys a wave equation
$\pt_{tt}w=c^2\pt_{xx}w$ and the boundary condition at the impact point is the continuity of the impulse $M\pt_{tt}w=-EA\pt_xw$ at $x = 0$ \cite{love}.  The resulting expression for the stress profile along the rod for $0<x<ct$ is then
\begin{eqnarray}
F(x,t)= E A \frac{U_0}{c} ~ \mathrm{exp}\left\lbrace -\frac{x}{L_p}\right\rbrace ~ \mathrm{exp}\left\lbrace \frac{ct}{L_p}\right\rbrace
\label{force}
\end{eqnarray}
where $L_p=M/(\rho A)$ is the length of the rod which has a mass equal to the projectile. Eq.~\ref{force} is valid until the front has reached the fixed end of the rod ($x=L$) \cite{note}. After that, for $t>L/c$, the wave rebounds, and the compression in the rod is the superposition of the incoming and reflected waves. The full solution of the reflected pulse problem can be solved with Saint-Venant's ``continuing equation", in which the solution $w(x,t)$ during the $n$th time range $n L/c < t < (n+1) L/c$ is obtained from the solution during the preceding time range \cite{love,goldsmith}. However, to a good approximation, the stress doubles after the first rebound.

Of crucial importance to the shape of the compression wave is the characteristic length $L_p$. Typically in our experiments $M \simeq 25$ g, which for our experiments gives $L_p=2-4$m, about an order of magnitude longer than the rod itself! Thus the exponential factors in Eq.~(\ref{force}) are of order unity, and the compression wave is a step pulse travelling at speed $c$, with magnitude

\begin{equation}
F_0 \simeq E A \frac{U_0}{c}.  
\label{force2}
\end{equation}

\begin{figure}[t]
\begin{center}
\includegraphics[width=8cm]{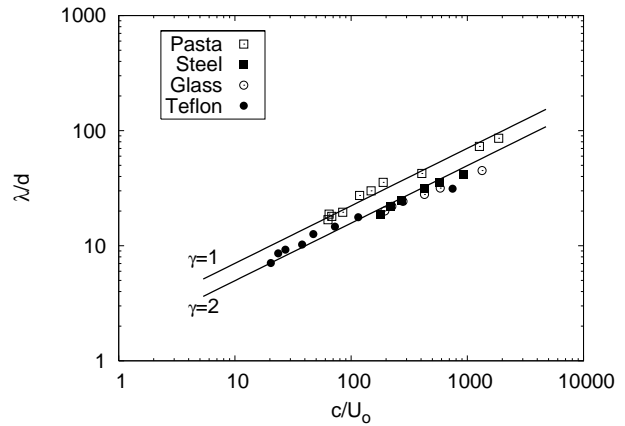}
\caption{The nondimensional buckling wavelength $\lambda / d$ vs the normalized impact speed. The $\gamma$ factor accounts for the reflection of the initial impulse (see text).  The lines correspond to the scaling law in \protect Eq.~(\ref{lawsscaled}), plotted for $\gamma = 1$ and 2.}
\label{wavelength}
\end{center}
\end{figure}

Inserting this into Eq.~(\ref{laws}) leads to the observed trends shown in Fig.~(\ref{speed}): $\lambda \sim d\sqrt{c/U_0}$, independent of $L$. This scaling law is valid if the time scale for the buckling $\tau_{\rm b}<L/c$ (the instability has to develop before the compression front has reached the rod end).  This occurs for low sound speed materials or large $L$.  However, if the pulse does reach the opposite end of the rod before the preferred mode develops, a reflected pulse is superimposed on the initial one, effectively doubling the stress. We therefore anticipate that the wavelength and growth rate will be given by

\begin{equation}\frac{\lambda}{d} = \frac{\pi}{\sqrt{2}} \sqrt{\frac{c}{\gamma U_0}}, \qquad {\rm and}\qquad \tau_{\rm b}^{-1} = \frac{2 \gamma U_0}{d}
\label{lawsscaled}
\end{equation}
with $\gamma = n+1$ for $n$ reflections. Using this scaling law to replot the wavelength data for all of the materials, diameters and impact speeds, we find excellent quantitative agreement over two decades (Fig.~\ref{wavelength}).  We find that all data align with curves with either $\gamma=1$ (no reflections) or $\gamma=2$ (one reflection), and no other adjustable parameters.

%%%%%%%%%%%%%%%%%%%%%%%%%%%%%%%%%%%%%%%%%%%%%%%%%%%%%%%%
%%%%%%%%%%%%%%%%%%%%%%%%%%%%%%%%%%%%%%%%%%%%%

High speed video also allows us to track the growth of the preferred buckling mode. For dry pasta ($d=1.1$ mm, $L=22$ cm, $\gamma=1$) with $U_0 = 3.5 \pm 0.1$ m/s, the above model predicts $\tau_{\rm b}=157 \pm 7~\mu$s. The initial increase of the buckling amplitude is indeed well fit by an exponential with a time constant of $151 \pm 8~\mu$s, consistent with the linear description of the instability development.

Note finally that Eq.~(\ref{force2}) can be understood heuristically as follows: at time $t$, the compressive wave front has travelled a distance $ct$. The compression of the rod is $\sim U_0 t$, which implies a compressive strain $\epsilon \sim U_0 t/ct = U_0/c$. From Hooke's law $F/A=E\epsilon$ (``{\it Ut tensio sic vis}''), we find $F\sim EAU_0/c$.

For brittle materials, dynamic buckling leads rapidly to the fragmentation of the rod. Because solid fragmentation typically results in a broad distribution of fragment sizes, with breaks occurring at apparently random locations, a statistical approach is usually taken \cite{mott47,grady85}.  The breakup of a rod under impact is known to give rise to broad distributions of fragment sizes \cite{ishii92,ching00}, with a highly skewed shape and a long tail, a phenomenology also encountered in other contexts, including liquid sprays \cite{lig}. However, the precise relation between the global statistics of fragmentation and the instabilities which produce it remains largely obscure.
 %odder93

As is evident from Fig.~1, the wavelength $\lambda$ determined by the dynamic buckling process leads to a preferred fragment length of $\lambda/2$, because the rod is likely to break at the points of highest deformation (maximum rod curvature). This $\lambda$-dependent breaking process emerges distinctly from a more random fragmentation. To investigate the interplay of these two processes, we studied the impact fragmentation of two different kinds of spaghetti, using the same projectile with $U_0 \simeq  3.5$ m/s (within 3\%). We shattered 300 pieces of angel hair pasta, where the length ranged from 21 - 22 cm and $d = 1.1$ mm  (within 2\%). After impact, we measured the length of each fragment larger than about two diameters, chosen in order to remain in the one-dimensional fracture regime, leading to a distribution of about 1200 fragments.  We also shattered 200 pieces of spaghetti,
with lengths from  22 - 23 cm and $d = 1.9$ mm  (within 2\%). This resulted in a distribution of about 1000 fragments.

\begin{figure}[t]
\begin{center}
\includegraphics[width=8cm]{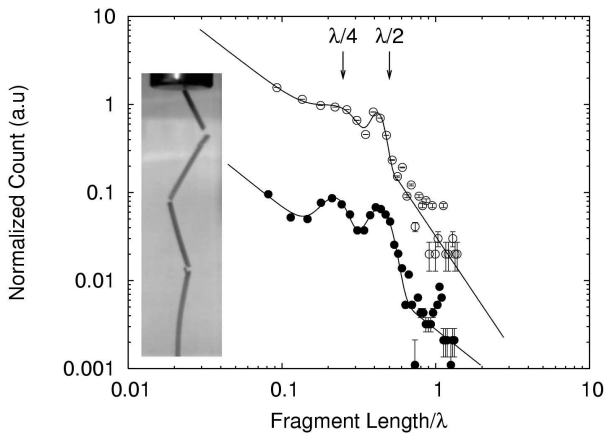}
\caption{\label{stats} Size distribution normalized by the experimental buckling wavelength $\lambda$ for impact fragmentation of pasta at $U_0 = 3.5$ m/s ($c \simeq 1400$ m/s): a) $d = 1.1$ mm, $\lambda=48$ mm), and b) $d=1.9$ mm, $\lambda = 70$ mm, vertically offset for clarity. The smooth lines are meant to guide the eye. The arrows indicate the fragment sizes $\lambda/4$ and $\lambda/2$. The inset shows a breaking event (captured at $\Delta t =$ 76  $\mu$s) contributing to both the $\lambda/2$ and $\lambda/4$ peaks.
}
\end{center}
\end{figure}

The experimental distributions are shown in Fig.~\ref{stats}. In both cases two broad peaks are discernable, indicating not one but two preferred lengths. These lengths are close to $\lambda/2$ and $\lambda/4$ (arrows in the figure). High speed video images show that the sinusoidal bending of the pasta often has a node at the projectile end, from which the distance to a maxiumum curvature point would be $\lambda/4$ (see inset Fig.~\ref{stats} and also Fig.~1c-d).  This would contribute one $\lambda/4$ fragment for each impact. The fact that both peaks are slightly lower than expected can be explained by concentrated breaking events near the ends of the main fragments, from which small pieces detach with much less rotation than the larger ones (see Fig.\ref{pasta}).

Digital video taken at much higher rates (20,000 fps) shows that the breaks do not occur simultaneously, but they are nearly so (within tens of microseconds, compared to the instability growth timsecale of about a hundred microseconds). This suggests that the fragment distribution is determined more by stress dynamics before breakup than by a sequential, multiplicative breaking often invoked in fragmentation models \cite{redner}, as first suggested by Kolmogorov \cite{kol}. However the precise rules governing the interaction between the primary compression wave, the possible rebounds at the rod extremities, and the buckled rod shape, remain to be discovered.

%%%%%%%%%%%%%%%%%%%%%%%%%%%%%%%%%%%%%%%%%%%%
%%%%%%%%%%%%%%%%%%%%%%%%%%%%%%%%%%%%%%%%%%%%

The Euler buckling of a slender rod is one of the simplest and most general instabilities of a solid, easily observable with almost any material. What we have shown here is that, using the ideas of Saint-Venant on elastic collisions, the dynamic buckling wavelength can be related to the impact velocity. Surprisingly, the scaling law for the buckling wavelength, which we have demonstrated in teflon, pasta, glass, and steel, depends on only two properties of the rod: the speed of sound and the diameter.  Moreover, the fragmentation of a solid under impact, usually conceived of as a random process to be treated statistically, has been shown here to include the imprint of the deterministic buckling process leading to breakup. Our simple experiments suggest that, by ``picking up the pieces", more complicated problems, such as magma fragmentation in explosive volcanic eruptions \cite{dingwell96} or the crushing patterns observed in carbon nanotubes under axial stress \cite{lourie98}, may still retain information, albeit in statistical form, on the initiating physical processes by which the impulse of impact is distributed throughout the solid before fragmentation.

We thank R.~Geist for experimental assistance and J.~Duplat for help in understanding Saint-Venant's solution.  AB acknowledges support from the National Science Foundation (CAREER grant DMR-0094167).

%\newpage


\begin{thebibliography}{99}
\bibitem{note} Eq.~(\ref{force2}) was also used by Hopkinson \cite{hopkinson} for an impulsively stretched wire, which does not buckle.
\end{thebibliography}
\end{document}